\documentclass[sigconf]{acmart}

\AtBeginDocument{%
  }

\copyrightyear{2026}
\acmYear{2026}
\setcopyright{cc}
\setcctype{by-nc-nd}
\acmConference[CHI EA '26]{Extended Abstracts of the 2026 CHI Conference on Human Factors in Computing Systems}{April 13--17, 2026}{Barcelona, Spain}
\acmBooktitle{Extended Abstracts of the 2026 CHI Conference on Human Factors in Computing Systems (CHI EA '26), April 13--17, 2026, Barcelona, Spain}
\acmDOI{10.1145/3772363.3798399}
\acmISBN{979-8-4007-2281-3/2026/04}

\begin{document}

\title{Workmanship of Learning: Embedding Craftsmanship Values in AI-Integrated Educational Tools}

\author{Tuan-Ting Huang}
\email{t.t.huang@tue.nl}
\orcid{0009-0001-3352-7357}
\affiliation{%
  \institution{Eindhoven University of Technology}
  \city{Eindhoven}
  \country{The Netherlands}
}

\author{Janet Yi-Ching Huang}
\affiliation{%
  \institution{Eindhoven University of Technology}
  \city{Eindhoven}
  \country{The Netherlands}
}
\email{y.c.huang@tue.nl}
\orcid{0000-0002-8204-4327}

\author{Stephan Wensveen}
\affiliation{%
  \institution{Eindhoven University of Technology}
  \city{Eindhoven}
  \country{The Netherlands}
}
\email{s.a.g.wensveen@tue.nl}
\orcid{0000-0001-8804-5366}

\renewcommand{\shortauthors}{Huang et al.}

\begin{abstract}

Generative AI's emphasis on automation and efficiency challenges design education, where learning is grounded in exploration, reflection, and responsibility. This work introduces AI Craftsmanship, a value-oriented framework drawing on craftsmanship traditions that emphasize risk, rhythm, and care as central to learning through making. Through a Research through Design (RtD) approach, we designed an AI-integrated creative coding tool embedding these values into interactions and interface rather than outcomes. The tool supports designers learning generative pattern-making with p5.js by constraining AI, encouraging iterative experimentation, and foregrounding reflection. We studied the tool with five design practitioners through one-hour sessions and semi-structured interviews. Findings show craft values manifest unevenly: risk and rhythm shape early sense-making, while care emerges through reflective practices. Emergent values---such as aesthetic judgment and confidence---also motivated learning. AI Craftsmanship mediates values, tools, and materials, offering a value-driven perspective on designing AI systems for reflective, responsible, craft-informed learning in design education.
\end{abstract}




\begin{teaserfigure}
  \centering
    \includegraphics[width=\textwidth]{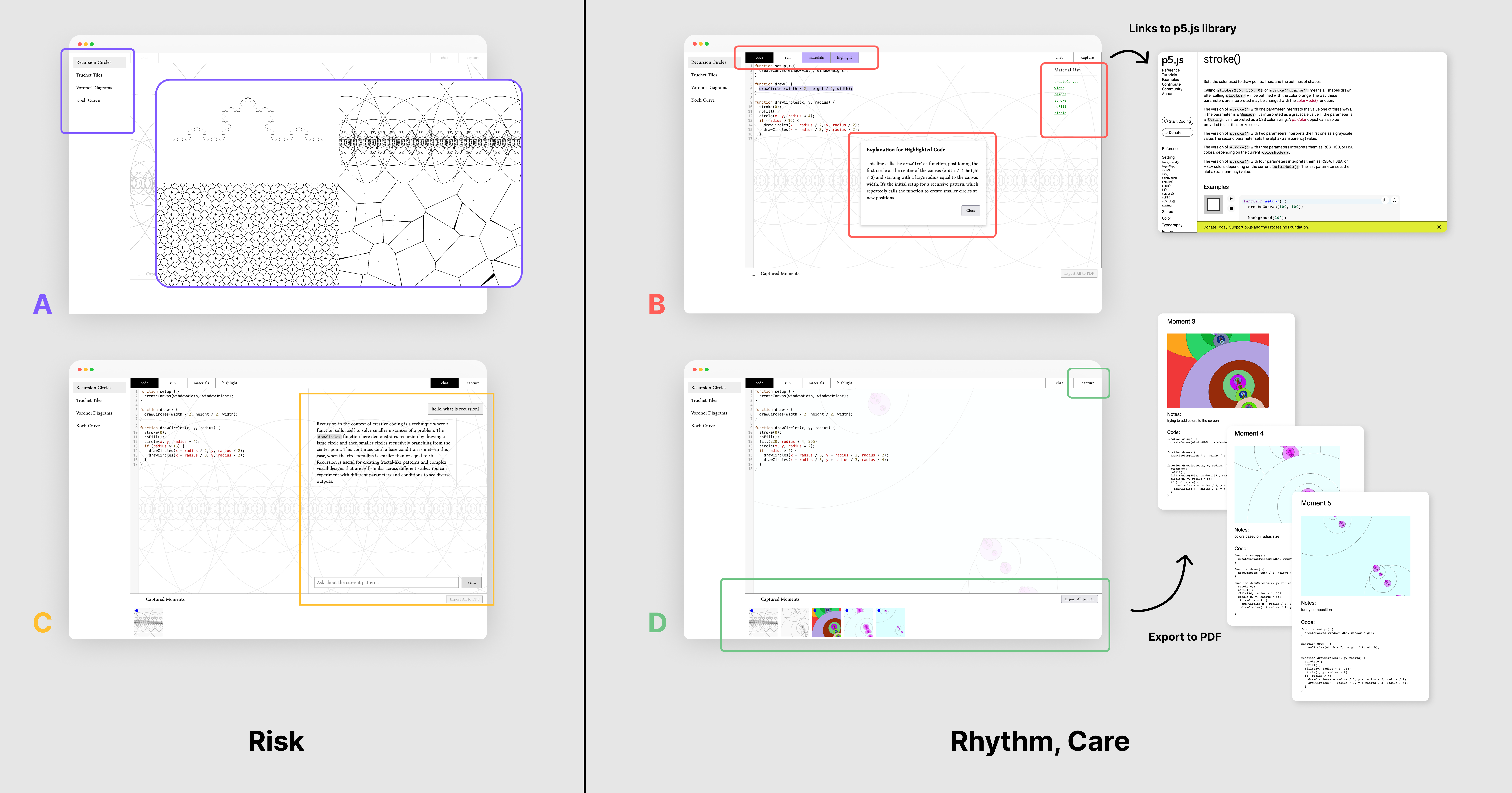}
    \caption {The tool designed in our research process as a means to elicit AI Craftsmanship values ---risk, rhythm, and care ---in interface and interaction design. Tool features include: (A) Pattern List, (B) Coding Toolbox, (C) Chat Interface, and (D) Capture. }
    \Description{A composite interface overview of an AI-supported creative coding tool. Four annotated screenshots are labeled A–D. (A) shows a pattern list and canvas displaying multiple generative geometric patterns. (B) shows a code editor with a coding toolbox, highlighted lines of code, an AI-generated explanation, and links to p5.js documentation. (C) shows a constrained chat interface where users ask conceptual questions about code. (D) shows a capture and reflection panel displaying a sequence of saved visual outputs, code snapshots, and notes that can be exported as a PDF. Labels indicate how features relate to the values of risk, rhythm, and care.}
    \label{fig:tool}
\end{teaserfigure}


\maketitle

\section{Introduction}
Craftsmanship has long held a central place in art and design history. Beyond form and aesthetics, it foregrounds judgment, responsibility, and attentiveness in the process of making. Craft-based pedagogy has been valued for integrating cognitive, practical, and affective dimensions of learning—what Cheatle and Jackson describe as the unity of head, hand, and heart \cite{cheatle2023recollecting}. Such traditions emphasize learning through material engagement, where uncertainty is sustained and decisions matter.

However, the rapid integration of generative AI into creative practice and design education unsettles this tradition. While AI systems often foreground speed, automation, and optimization, these priorities risk displacing slower, exploratory learning processes through which learners develop judgment, agency, and responsibility. As AI increasingly generates code, suggests compositions, or completes designs on students’ behalf, a critical question emerges: how can design education cultivate care, responsibility, and critical engagement when uncertainty is minimized and decision-making are externalized?


Craftsmanship literature offers a productive lens for examining this challenge by centering \textit{how} work is done and \textit{how} practice shapes the practitioner. Pye  characterizes craftsmanship as the ``workmanship of risk'' \cite{pye1968nature}, where quality is continually at stake during making. Ingold emphasizes rhythm as attentiveness developed through repetition and adjustment~\cite{Ingold2006Walking}, while Sennett frames crafting as an ethical commitment to care~\cite{sennett2008theC}. Together, these perspectives highlight craft values---risk, rhythm, care---that cultivate agency, attention, and responsibility through making.

However, these craft values have not yet been substantively engaged in AI-supported design learning. Research on AI in education has largely focused on efficiency, scaffolding, and assessment~\cite{Wang:2024}, while work on creative AI emphasizes authorship and co-creativity \cite{lawton2023drawing}. Less attention has been paid to how AI tools might be designed to support value-oriented dimensions central to craft pedagogy: sustaining productive uncertainty rather than automating problem-solving, cultivating epistemic judgment rooted in values and tacit understanding, and fostering care and attentiveness to process rather than optimizing for task efficiency. If craft values are to remain relevant in AI-mediated contexts to prioritize process in learning, they must be actively operationalized into tool design rather than assumed to transfer from traditional making practices.

In response, this paper proposes \textit{AI Craftsmanship} as a value-oriented framing for AI-supported design education. Rather than assigning AI a fixed role, AI Craftsmanship focuses on how AI mediates relationships between learners, tools, and materials through interaction. We draw on three craft values---risk, rhythm, and care---as an initial lens for exploring how AI systems can support learning without collapsing uncertainty or displacing responsibility. These values are not positioned as exhaustive; instead, they serve as an entry point for examining how craft values can be surfaced, tested, and expanded through the design of AI-integrated learning environments. 

We selected these three values from craftsmanship literature for their particular relevance to contemporary AI contexts: risk in relation to algorithmic uncertainty and indeterminacy; rhythm in sustaining engagement within iterative making processes; and care in addressing questions of authorship, ownership, and responsibility that become increasingly blurred as AI participates in creative work.

We investigate this framing through a Research through Design approach, developing an AI-integrated creative coding tool within a bachelor's course, \textit{Digital Craftsmanship}, in the Industrial Design department at Eindhoven University of Technology. The tool supports novice designers learning generative pattern-making with p5.js and constrains the tendency of AI toward efficiency and completion, encouraging iterative experimentation and reflection. We studied the tool with five designers and students through one-hour making sessions and semi-structured interviews.

This work contributes (1) AI Craftsmanship as a conceptual framework positioning AI as a mediating layer between values, tools, and materials, and (2) an RtD account of operationalizing craft values in an AI-supported learning tool, revealing both the possibilities and limits of this approach for supporting reflective, responsible design learning.

\section{Related Work}
\subsection{Craftsmanship and digital craftsmanship}
Craft emerges from the ongoing negotiation between practitioner, tool, and material. Ingold  emphasized attentiveness within rhythmic flow \cite{Ingold2006Walking}, where each action is adjusted in response to material and circumstance. Such rhythm is not automation but situated attention—a cadence that develops through experience and context. Judgement and skills emerged through these cycles of interactions. Ultimately, these iterative cycles are to serve the dedication of the result. A craftsperson embodies dedication and responsibility driven by the care of the material and the quality of their output \cite{sennett2008theC}. On the other hand, Pye's focus of the workmanship of risk characterizes craft as a practice where outcomes are uncertain and quality depends on the maker's judgment \cite{pye1968nature}. This contrasts with the workmanship of certainty, where automated processes produce predictable results. Craft techniques develop through a dialectic between knowing established methods and experimenting through error \cite{sennett2008theC}, cultivating agency and responsibility through practice. 

As design practice entered digital contexts, scholars extended these ideas through digital craftsmanship \cite{McCullough1996AbstractingCT}. Kolko argues for craftsmanship values---attention, iteration, judgment---in software and interaction design \cite{kolko2011craftsmanship}, while some emphasize rich material engagement in interactive artifacts. Work on electronic crafts and hybrid fabrication further explores expressive digital materials that support situated making \cite{brandt2016laserAM}. While these perspectives reaffirm craftsmanship values in computational contexts, the emergence of AI introduces new forms of agency, opacity, and uncertainty that challenge existing framings and call for renewed attention to how craftsmanship values are enacted in AI-mediated design practice.

\subsection{Creativity support tools}
A creativity support tool is any tool that can be used by people in the open-ended creation of new artifacts. It can span across multiple domains and activities, from music making, creative writing, to image and video editing \cite{CherryLatulipe2014CSI}. Research on creativity support tools (CSTs) emphasizes scaffolding exploration, iteration, and reflection rather than accelerating production \cite{shneiderman2007cst}. Recent syntheses highlight increasing attention to supporting creative processes over time, particularly as AI enters these environments \cite{frich2019mappingCST}. Work on AI and co-creativity explores systems that collaborate with, rather than replace, human creators \cite{lubart2005howCC}. 
Despite CST attention to the process of creation, it focuses mostly on supporting creativity of the users. On the other hand, constructionist approaches frame learning as emerging through making and revising artifacts \cite{papert1980mindstorms}, to involve learners in “reflecting on more complex aspects of their own thinking”. Our work builds on CST traditions but shifts emphasis toward constructionist learning to incorporate more craftsmanship-informed values. Through learning-through-making and reflection-in-action \cite{schon1983reflective}, we want to foreground ethical values such as risk, rhythm, and care as central to how learners engage with AI in design education.

\section{Methodology}
This study adopts a Research through Design (RtD) methodology \cite{zimmerman2007research,gaver2012what}, combining first-person reflective research \cite{desjardins2021introduction, neustaedter2012autobiographical} grounded in literature, tool design and development, and a qualitative empirical study. Prior to this research, the first author had two years of experience in creative coding classrooms, including supporting open-source libraries such as p5.js and Processing, designing learning tools, and contributing to educational initiatives in creative coding. This sustained engagement with learners---attending closely to moments of struggle, pacing, and creative decision-making---cultivated a practical sensitivity to how learners encounter uncertainty and develop their own making practices in coding contexts, the same values this research foregrounds. These first-hand experiences informed the design of the tool through iterative, first-person reflection. The tool and study together functioned as a site of inquiry, surfacing and testing assumptions about how craftsmanship values can be translated into interaction design. This work explores how these values manifest in practice, where tensions arise between design intentions and learner experience, and what additional values emerge through use.

Rather than producing a finished artifact, we designed a conditional environment---one that shapes the circumstances for learning without determining outcomes---intended to elicit the values of risk, rhythm, and care. Because values cannot be directly embedded into a system, we operationalized them through interface and interaction design choices that prioritize engagement with process over outcome, foregrounding \textit{how} making unfolds rather than \textit{what} is produced. Based on these considerations, we intentionally constrained the role and capabilities of AI while providing calibrated scaffolding that preserves space for exploration. These constraints aim to maintain uncertainty while offering sufficient guidance to maintain the learners' flow, allowing participants to remain within a productive space of constructive struggle and ``not knowing.'' In doing so, the tool resists speed, automation, and auto-completion, instead encouraging slower interaction, exploration, and reflective engagement as central aspects of learning with AI.

\subsection{Tool features and design}
\label{subsec:tool-feature}
To operationalize these design considerations, we developed the tool to support learning creative coding in p5.js through generative patterns and AI-assisted interaction. We selected p5.js because it is an open-source library designed to be accessible to beginners while supporting creative exploration. The tool is designed for designers with little or no coding experience, positioning AI as a mediator of learning rather than an automated solution generator. Each feature emerged through iterative design and first-person testing, shaped by the central design consideration: how might this specific interaction elicit---rather than shortcut---craftsmanship values?

\subsubsection{Pattern list (risk)}
Users begin with a curated set of generative pattern templates that provide orientation without prescribing outcomes (Fig.~\ref{fig:tool}(A)). Drawing on instructors' experience in the situated course, we observed that novice coders often find a blank canvas intimidating and benefit from initial scaffolding. However, too much structure risks eliminating the productive struggle central to craft learning. To balance support and openness, the provided patterns are beginner-friendly and remain close to their underlying algorithms, exposing core structural logic rather than polished results. Templates also create an intentional mismatch with the’ expected visual outcomes anticipated by learners, creating space for interpretation and productive struggle. Rather than starting from an empty editor, learners modify, break, and reinterpret existing structures, sustaining uncertainty while supporting personal expression. In this way, the pattern list frames risk as an ongoing condition of making rather than something to be eliminated at the outset.

\subsubsection{Coding toolbox (rhythm, care)}
The coding toolbox supports multiple levels of code engagement, including running sketches, inspecting referenced p5.js functions through official interactive documentation, and highlighting code for AI-generated explanations (Fig.~\ref{fig:tool}(B)). These features are designed to maintain the flow of learning by keeping attention anchored within the coding environment. By allowing learners to move fluidly between execution, reference, and explanation, the toolbox supports iterative experimentation (rhythm) and close, situated attention to code as material (care). Rather than relying on opaque automation, learners directly examine how specific code structures operate, encouraging deliberate adjustment and reflective engagement over time.

\subsubsection{Constrained AI chat interface (risk)}
Constrained conversational AI supports conceptual questions, debugging, and ideation without generating full solutions (Fig.~\ref{fig:tool}(C)). Early versions of the tool used an unconstrained GPT-5.0 API, which resulted in learners frequently copying and pasting AI-generated code with minimal reflection. To counteract this, we iteratively redesigned the system prompt to restrict response length and shift AI's role from problem-solving to sense-making. Instead of producing complete answers, the AI responds with conceptual guidance and reflective questions. By limiting certainty and verbosity, the chat interface scaffolds learning while preserving productive struggle, sustaining learner agency, and maintaining risk as a generative condition within the creative process.

\subsubsection{Capture and reflection (rhythm, care)}
The Capture feature records snapshots of code and visual output alongside user annotations, allowing learners to document and reflect on their process over time (Fig.~\ref{fig:tool}(D)). We designed annotation as both documentation and attentiveness, foregrounding code as material and making shifts in understanding visible across iterations. Reviewing these captured moments supports temporal awareness of the making process, encouraging continued exploration and reframing outcomes as moments within an unfolding practice rather than final results.

Throughout development, the tool served as a reflexive research probe. Iterative design decisions and early tests surfaced how craftsmanship values could be negotiated through interface design, informing both the final tool and the subsequent analysis.

\subsection{Study design}
\label{subsec:study-design}
This work surfaced assumptions about how craftsmanship values might be translated into interaction, but it could not reveal how learners would actually engage with these design. To extend the inquiry, we conducted a preliminary qualitative study with five design practitioners to examine how learners engaged with our tool. 

The study was conducted in the Netherlands, where all participants resided and were recruited. Participants were sampled to reflect varied design backgrounds and levels of coding experience. The participants included a master’s student with a craft and graphic background; a recent master’s graduate in industrial and social design; two design researchers working in AI and HCI; and a professional graphic designer with over 15 years of experience. Two participants reported minimal prior exposure to creative coding and identified as novices; the remaining three had no prior experience. None had previously used the tool.

Each study session lasted approximately one hour: a background questionnaire (5 min), a guided walkthrough of the tool (5 min), an independent making session responding to a design prompt---create a generative pattern for a community farmer's market day (30 min) and a semi-structured interview (20 min). 

We collected interview transcripts, participants' captured snapshots and annotations, and observation notes. Captured snapshots served as elicitation materials during interviews, prompting reflection on specific moments. Data were analyzed using reflexive thematic analysis~\cite{Braun:2019}, with codes generated inductively and organized around participants’ learning experiences, interactions with AI, and perceptions of process and uncertainty. This study provided empirical insights into how the design values manifested in practice and how participants made sense of AI as part of their creative workflows. We synthesize and discuss our findings on three focuses: how values manifested, where tensions arose, and what additional values emerged in the next section.

\section{Findings and Discussion}
\subsection{Revisiting craft-informed values: risk, rhythm, and care}
Our findings show that risk, rhythm, and care do not emerge uniformly but unfold over time, shaped by how AI support is constrained and framed. Tensions arose primarily around the pacing of AI support and the perceived completeness of AI outputs.
\textbf{Risk} surfaced most visibly in interactions with the Chat feature. Participants described AI responses as ``magical,'' producing outputs that felt overly complete and risked diminishing productive struggle. By contrast, when participants struggled with the abstract materiality of code, the Highlight feature positions AI as a translator of computational materiality---making rules and structures visible without resolving uncertainty. In this role, AI helped learners remain within a constructive zone of not-knowing.
\textbf{Rhythm} emerged as a temporal learning trajectory rather than a steady flow. Early interactions focused on orientation and rapid parameter changes aimed at sense-making. Care and intentionality appeared only after familiarity developed. Participants expressed a desire for a gradual transfer of agency away from AI, suggesting that supportive AI tools should choreograph pacing rather than optimize early performance.
\textbf{Care} was least visible within the short study duration but appeared through early documentation and reflection practices. Participants who used the Capture feature described it as supporting attentiveness to decisions and process, indicating that care may first emerge as process-oriented awareness rather than outcome refinement. Longer engagement would likely reveal more developed forms of care, including responsibility toward the artifact and critical evaluation of AI contributions, which should be explored in the future.
Importantly, we position risk, rhythm, and care as an initial scaffold, not a closed set of values. The ongoing articulation and expansion of values through tool use is itself central to the spirit of AI Craftsmanship.

\subsection{Emergent values beyond the initial framework}
Additional values emerged empirically through use. \textbf{Security and confidence} functioned as enabling conditions for risk-taking: participants evaluated success relative to experience rather than polish. \textbf{Aesthetic judgment} emerged as a primary driver of engagement and sense-making, with creativity described as present in ``every parameter change.'' These values sustained motivation during early encounters with computational material and shaped learning trajectories. These findings reaffirm craftsmanship as a tacit, reflective practice where learning emerges through making. Designing AI-supported learning environments must therefore attend not only to ethical values but also to experiential conditions---like confidence and aesthetic judgment---that sustain exploration.

\subsection{AI Craftsmanship as a mediating layer between values, tools, and materials}
Taken together, our findings frame AI Craftsmanship as a mediating layer between values, tools, and materials. AI neither functions solely as an efficiency tool nor as an autonomous creative agent; instead, it shapes how learners encounter uncertainty, pace engagement, and develop responsibility toward outcomes. In our tool, this mediation was visible in how AI shifted roles across the learning process. During early sense-making, AI served as a translator---rendering computational logic visible visible without collapsing uncertainty. During debugging, it functioned as a thought-partner---offering scaffolding that preserved learner agency. During explanation and aesthetic negotiation, it became a catalyst---generating possibilities that prompted learners to clarify their intent while retaining judgment over outcomes. These shifting roles suggest that AI's function in learning environments is not fixed but situationally enacted through interaction. This framing moves beyond a strict tool-material dichotomy. AI Craftsmanship embraces the ambiguity of AI's roles and offers language for intentionally designing and reflecting on these transitions. For tool designers, it provides a value-oriented lens for interaction design. For learners, it shapes experiential qualities of engagement---embedding ethical responsibility through use rather than abstraction.

\begin{figure}
    \centering
    \includegraphics[width=1\linewidth]{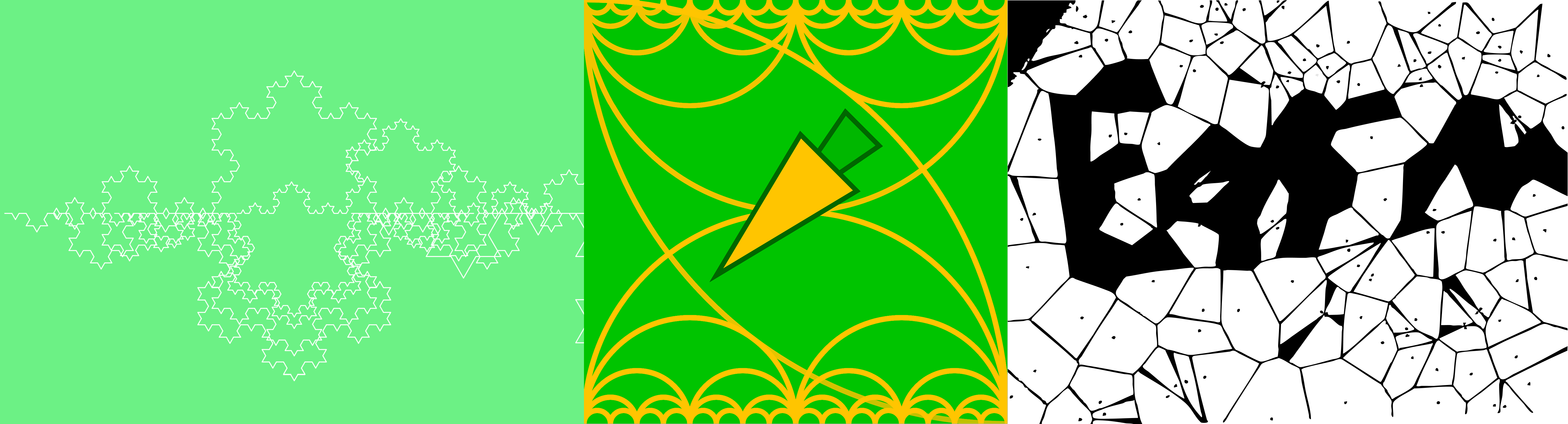}
    \vspace{-5mm} 
    \caption{Example outcomes for the qualitative study with participants.}
    \Description{Three generative visual outcomes displayed side by side. Left: a light green background with a white, recursive, snowflake-like pattern formed by repeated geometric outlines. Center: a bright green background with layered yellow circular arcs and an angular, overlapping triangular form in the foreground. Right: a black-and-white composition resembling a cracked or cellular surface, with irregular polygonal shapes and scattered points creating a fragmented texture.}
    \label{fig:outcome}
    \vspace{-3mm} 
\end{figure}

\section{Conclusion and Future Work}
This work proposed AI Craftsmanship as a value-oriented framework for AI-supported design education. Through a RtD process, we developed an AI-integrated creative coding tool and examined how risk, rhythm, and care manifest in practice. Our findings reveal that  craftsmanship values unfold temporally and situationally---risk and rhythm shape early sense-making, while care emerges through reflective practices. We also identified emergent values---aesthetic judgment and confidence---that shape motivation and warrant attention alongside ethical considerations. We frame AI Craftsmanship as a mediating layer between values, tools, and materials---shaping how learners encounter uncertainty, pace engagement, and develop responsibility. Craftsmanship reminds us that learning is not only about acquiring skills but forming relationships with materials, technologies, and one's own making. Future work will explore sustained engagements and extend this framework to other creative learning contexts.

\begin{acks}
This research was carried out as part of the alignAI project and, as such, has received funding from the European Union's Horizon Europe research and innovation programme under the Marie Skłodowska-Curie grant agreement No. 101169473. We also sincerely thank all the participants, and the support and supervision from the Industrial Design Department at Eindhoven University of Technology.
\end{acks}

\bibliographystyle{ACM-Reference-Format}
\bibliography{reference}

\appendix

\end{document}